\documentclass[a4paper,fleqn,usenatbib]{mnras}

\usepackage{newtxtext,newtxmath}
\usepackage[T1]{fontenc}
\usepackage{ae,aecompl}

\usepackage{graphicx}	% Including figure files
\usepackage{amsmath}	% Advanced maths commands
\usepackage{amssymb}	% Extra maths symbols

\title[ ]{High-resolution spectroscopy of Boyajian's star during optical dimming events}
\author[M. J. Mart\' inez Gonz\'alez et al.]{
M. J. Mart\' inez Gonz\'alez$^{1,2}$\thanks{E-mail: marian@iac.es},
C. Gonz\'alez-Fern\'andez$^{3}$,
A. Asensio Ramos$^{1,2}$,
\newauthor
H. Socas Navarro$^{1,2}$,
C. Westendorp Plaza$^{1,2}$,
T. S. Boyajian$^{4}$,
J. T. Wright$^{5,6}$,
\newauthor
A. Collier Cameron$^{7}$,
J. Gonz\'alez Hern\'andez,$^{1,2}$
G. Holgado,$^{1,2}$
G. M. Kennedy,$^{8}$
\newauthor
T. Masseron$^{1,2}$
E. Molinari,$^{9}$
J. Saario,$^{10,11}$
S. Sim\'on-D\' iaz,$^{1,2}$
and B. Toledo-Padr\'on,$^{1,2}$
\\
$^{1}$Instituto de Astrof\' isica de Canarias, C/V\'{\i}a L\'actea s/n, E-38205 La Laguna, Tenerife, Spain\\
$^{2}$Departamento de Astrof\' isica, Universidad de La Laguna, E-38206 La Laguna, Tenerife, Spain\\
$^{3}$Institute of Astronomy, Madingley Road, Cambridge CB3 0HA, UK\\
$^{4}$Department of Physics and Astronomy, Louisiana State University, Baton Rouge, LA 70803 USA\\
$^{5}$Department of Astronomy \& Astrophysics, The Pennsylvania State University, 525 Davey Lab, University Park, PA 16802, USA\\
$^{6}$Center for Exoplanets and Habitable Worlds, The Pennsylvania State University, 525 Davey Lab, University Park, PA 16802, USA \\
$^{7}$Centre for Exoplanet Science, SUPA, School of Physics and Astronomy, University of St Andrews, North Haugh, St Andrews KY16 9SS, UK\\
$^{8}$Department of Physics, University of Warwick, Gibbet Hill Road, Coventry, CV4 7AL, UK\\
$^{9}$Telescopio Nazionale Galileo - FGG INAF, Bre\~na Baja, Tenerife, E-38712, Spain\\
$^{10}$Koninklijke Sterrenwacht van Belgi\"e, Ringlaan 3, B-1180 Brussels, Belgium\\
$^{11}$Institute of Astronomy, KU Leuven, Celestijnenlaan 200D, B-3001 Leuven, Belgium
}

\date{Accepted XXX. Received YYY; in original form ZZZ}

\pubyear{2018}

\begin{document}
\label{firstpage}
\pagerange{\pageref{firstpage}--\pageref{lastpage}}
\maketitle

\begin{abstract}

Boyajian's star is an apparently normal main sequence F-type star with a very unusual light curve. The dipping activity of the star, discovered during the  \emph{Kepler} mission, presents deep, asymmetric, and aperiodic events. Here we present high resolution spectroscopic follow-up during some dimming events recorded post-\emph{\emph{Kepler}} observations, from ground-based telescopes. We analise data from the HERMES, HARPS-N and FIES spectrographs to characterise the stellar atmosphere and to put some constraints on the hypotheses that have appeared in the literature concerning the occulting elements. The star's magnetism, if existing, is not extreme. The spots on the surface, if present, would occupy 0.02\% of the area, at most. The chromosphere, irrespective of the epoch of observation, is hotter than the values expected from radiative equilibrium, meaning that the star has some degree of activity. We find no clear evidence of the interstellar medium nor exocoments being responsible for the dimmings of the light curve. However, we detect at 1-2 sigma level, a decrease of the radial velocity of the star during the first dip recorded after the \emph{\emph{Kepler}} observations. We claim the presence of an optically thick object with likely inclined and high impact parameter orbits that produces the observed Rossiter-McLaughlin effect. 

\end{abstract}


%%%%%%%%%%%%%%%%%%%%%%%%%%%%%%%%%%%%%%%%%%%%%%%%%%

%%%%%%%%%%%%%%%%% BODY OF PAPER %%%%%%%%%%%%%%%%%%

%\section{April 2009 - May 2013. The misterious dipping activity of the Boyajian star.}
\section{The misterious dipping activity of Boyajian's star.}

KIC 8462852, recently also known as Boyajian's star, is a fascinating object that has attracted considerable attention, not only from the scientific community but also the general public. It is an apparently normal main-sequence F type star except for a very complex and unexpected photometric evolution. In \citet{BoyajianWTF}, this star was highlighted by the citizen project "Planet Hunters" as having a very peculiar light curve in the form of deep irregular dips as seen in the \emph{Kepler mission} photometry recorded during 4 years \citep[see Fig. 1 in][]{BoyajianWTF}. The main events were strong (15-20\%) asymmetric dimmings that lasted for a few days and were separated by two years. Some other 1-5\% dips appear scattered in the time series with no clear periodicity. 

Other stars in which such irregular light curves have been detected are, e.g. young or pre-main-sequence stars that have debris or proto-planetary discs \citep[e.g.][]{Scaringi2016} or possibly stars with ringed planetary companions \citep[e.g.][]{Kenworthy&Mamajek2015}. Intrinsic R-Coronae-Borealis-like activity (formation of clouds that obscure the stellar atmosphere) can also create such irregular light curves. On the older side of the age spectrum, the white dwarf WD 1145+017 shows a variability that is thought to be caused by tidally disrupted planetesimals producing similar complex lightcurves \citep[e.g.][]{Rappaport2016}. 

However, none of these scenarios were favoured by the observations presented in \citet{BoyajianWTF}. The star was classified as a main-sequence star with high probability. This means that the R-Coronae-Boralis activity is very unlikely since it is typical of G to F evolved supergiants. Also, a close debris or a proto-planetary disc is very unlikely around a main-sequence star. This fact was supported by the lack of any infrared excess, that would provide evidence for hot circumstellar material, as pointed out in \citet{BoyajianWTF, Marengoetal.2015} and further confirmed at mm and sub-mm wavelengths by \citet{Thompson2016}. 

In the original paper by \cite{BoyajianWTF}, the authors suggested that the irregular dips of the star may be due to the passage of a family of exocomets or planetesimal fragments resulting from a previous break-up event. These possibilities have been further explored by \citet{Bodman&Quillen2016}, \citet{NeslusanBudaj2017} and \citet{Wyattetal2018}. Other possible scenarios have been proposed, as reviewed in \citet{Wright2016} where they find inhomogeneities in the interstellar medium as the most plausible one.

In May 2017, four years after the end of the \emph{Kepler} observations, the community was alerted by the team of Dr. Boyajian of a new dipping event of the star. This very first observed dimming since the discovery of the star's behaviour in the \emph{Kepler} data was named \textit{Elsie} \citep{BoyajianTEAM2018}. Since then, the star has suffered from several other dimming events, and several names have been adopted for each dip by the project's Kickstarter supporters\footnote{https://www.kickstarter.com/projects/608159144/the-most-mysterious-star-in-the-galaxy}. Up to now, the events have been aperiodic, and have obscured the stellar flux by 5\%, at most. The most prominent dips happened in March 2018, and hence were not published in \cite{BoyajianTEAM2018}\footnote{The photometric follow-up of the star to date is publicly available in http://www.wherestheflux.com/blog}. Some new tentative theoretical explanations were prompted by the dimming events in May 2017 which include multiple dust-enshrouded objects \citep{Neslusan&Budaj2017}, a ringed planet with Trojan asteroids \citep{Ballesterosetal.2017}, clumpy material in the outer Solar System \citep{Katz2017}, long-term variability after the consumption of a planet \citep{Metzgeretal.2017}, or intrinsic convective inhibition by the star itself \citep{Foukal2017}. 

Interestingly, using ground-based instrumentation, \cite{BoyajianTEAM2018} and \cite{Deegetal.2018} were able to perform colour photometry and spectrophotometry, respectively, and study the chromatic properties of the dips. Both works favour the presence of circumstellar dust occulting the star since the blue wavelengths were more absorbed than the red ones during the dips.

Apart from the irregular and aperiodic dips, the light curve of the star presents some low-amplitude, periodic modulation. A Fourier analysis of the light curve performed in \cite{BoyajianWTF} revealed a prominent modulation feature with a period of 0.88 days that is present through all the light curve, though with variable amplitude (0.1\% variation on average). The authors claim that the 0.88 day period is very likely associated to the rotation of the star due to the passage of inhomogeneities in the surface such as spots. Though the presence of a very close companion (real or just a projection effect) can not be ruled out as the origin of this periodicity \citet{Makarov&Goldin2016}, recent observations by \citet{Clemensetal.2018} discard that this faint companion and Boyajian's star form a bound system.

Doppler tomography should reveal if the surface is covered by contrasted flux regions (e.g. cold spots) that are modulated with this 0.88 day rotation. This would be extremely interesting as there are very few examples of active F stars \citep{Martinez-Arnaizetal.2010} and those detected exhibit quite a complex behaviour \citep{Mathuretal.2014}. The surface tomography using the Doppler imaging technique is one of the objectives of the present paper. This tomography and the possible magnetic activity of the star are discussed in section 2. Serendipitously, another goal of this paper has been the study of the spectral variability of the star during the dimming events. Spectroscopy in and out of a dip can help us reject or strengthen some of the scenarios proposed to explain obscurations events in the star. This study will be presented in sections 3 and 4.

%\section{starting on May 2017. The Boyajian star is dipping again.}

\section{Boyajian's star is dipping again. High-resolution spectroscopic follow-up.}
\label{dipping}

On 11-18 May 2017 we were at the Mercator telescope at the Observatorio del Roque de Los Muchachos to perform high resolution spectroscopy of Boyajian's star. The initial goal was to study the possible magnetic activity in the star during its quiescent state to relate it to the observed 0.88 day periodicity. Luckily, and unexpectedly, the star started dipping on May 18. Given the special circumstances, we could extend our observational period, allowing unprecedented high-resolution spectroscopic monitoring of a full dipping event of Boyajian's star. 

\begin{figure}
\includegraphics[width = 0.8\columnwidth, angle = -90]{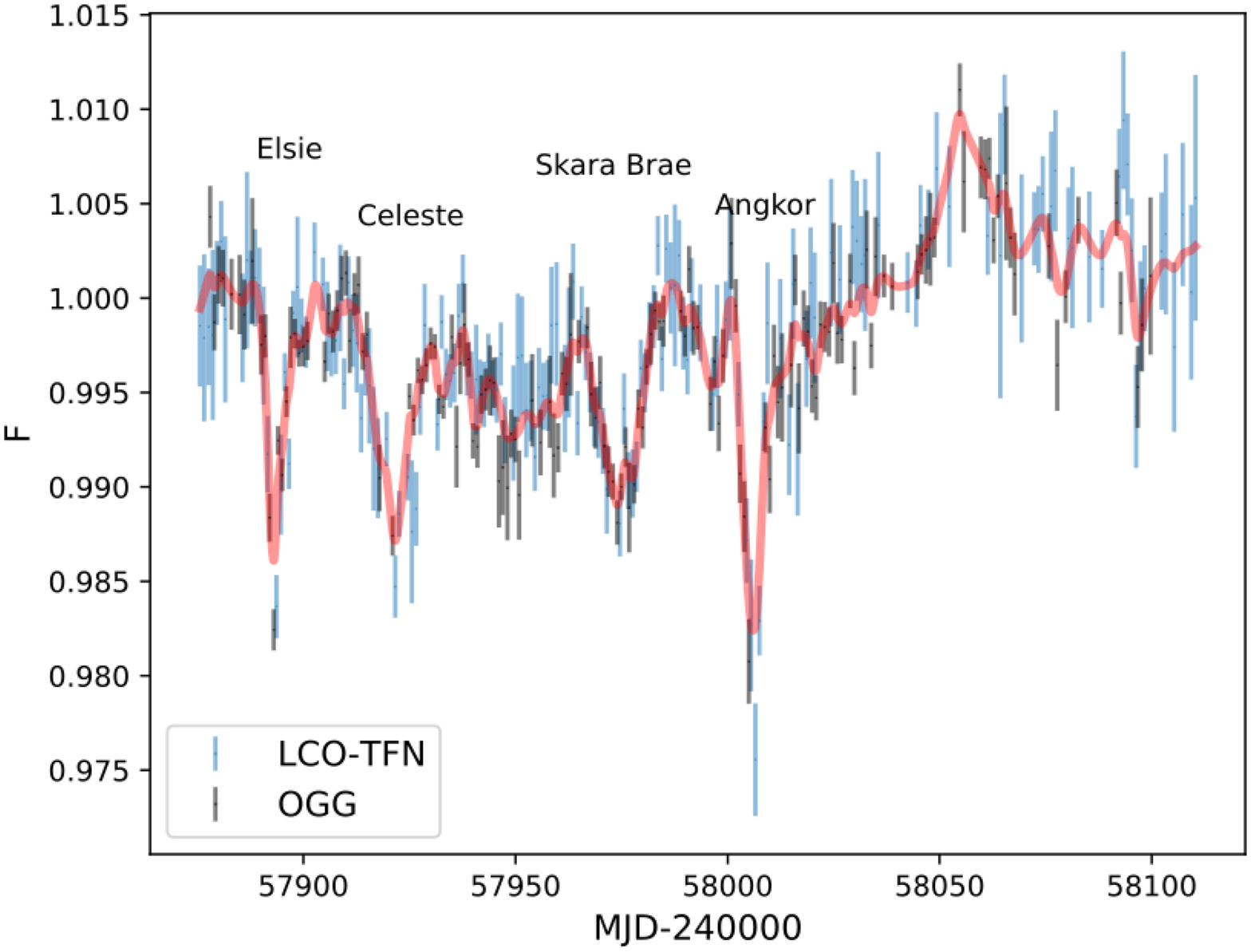}
\caption{Evolution of the flux for Boyajian's star during the dipping episodes described in \citet{BoyajianTEAM2018}, as measured from Tenerife. The thick red line is a Gaussian process interpolation of the data, that will be used as a timeline for all the plots in this work.}
\label{phot_curve}
\end{figure}

To contextualise our observations and results, Fig. \ref{phot_curve} displays the evolution of the flux during the dipping episodes described in \citet{BoyajianTEAM2018}. The first dip after the \emph{Kepler} observations was named \emph{Elsie}. During \emph{Elsie}, the star dimmed by less than 2\%, had an abrupt ingress, a very narrow dip, and a more extended egress. The second one, \emph{Celeste} was similar to \emph{Elsie}. After Celeste, there was a period with lower flux but no clear dip. \emph{Skara Brae} was a more symmetric dip with, possibly, a dramatic drop of the flux to 3\% in the middle. The last one, \emph{Angkor} had a shape similar to \emph{Elsie} but was the deepest of this period, dimming the star by more than 2\%. Note in the Figure that there is not a clear basal flux during the whole period. Here, we provide high-resolution spectroscopy throughout the \emph{Elsie} event,
after the \emph{Celeste} one, and within the \emph{Angkor} dip. 

\begin{table*}
\begin{tabular}{llllll}
{\bf Instrument} & {\bf MJD period} & {\bf \# of spectra} & {\bf t$_\mathrm{exp}$ [s]} & {\bf spectral resolution}\\
\hline
HERMES & 57885.6 - 57891.7 & 32 & 1800 & 85000\\
HERMES & 57893.6 & 3 & 1800 & 85000 \\
HERMES & 57896.6 & 3 & 1800 & 85000\\
HERMES & 57899.6 & 6 & 1800 & 85000\\
HERMES & 57941.4 - 57942.6 & 25 & 1800 & 85000\\
HERMES & 57944.4 - 57947.7 & 46 & 1800 & 85000\\
FIES & 57898.0 & 6 & 1000 & 46000\\
FIES & 58007.0 & 4 & 1530 & 67000 \\
FIES & 58012 & 2 & 1530 & 67000 \\
FIES & 58014.1 & 2 & 1530 & 67000\\
FIES & 58025.9 & 2 & 1530 & 67000 \\
FIES & 58029.9 & 1 & 1800 & 25000\\
FIES & 58034.8 & 1 & 1200 & 25000\\
FIES & 58036.9 & 3 & 1530 [2] \& 1200 [1] & 67000 [2] \& 25000 [1]\\
HARPS-N & 57894.2 & 1 & 2700 & 115000\\
HARPS-N & 58007.0 & 1 & 900 & 115000\\
HARPS-N & 58008.0 & 1 & 900 & 115000\\
HARPS-N & 58056.8 & 1 & 2700 & 115000
\end{tabular}
\caption{Summary of spectroscopic observations of Boyajian's star taken during May - October 2017.}
\label{tabla}
\end{table*}

The spectroscopy of the \emph{Elsie} event and after \emph{Celeste} was performed with the HERMES instrument \citep{hermes}, a fibre-fed spectrograph attached to the Mercator telescope. We recorded the spectral range between 377 to 900 nm with a resolving power R=85000. We took continuous exposures of 30 min during the whole night (see Table \ref{tabla} for more details). During the May campaign, we lost some nights because of the weather conditions as well as for technical problems. Overall, we could cover the full \emph{Elsie} event: 20 spectra pre-dip, 18 within the dip, and 6 after it. In July, we obtained 71 spectra that allowed us to conveniently map in phase the 0.88 day variability (to sample the full surface we spent almost 7 rotations of the star, assuming a 0.88 day period). The signal to noise (SN) ratio around H$_\alpha$ of individual spectra is 14, on average. However, in May's campaign, the weather conditions were not optimal and highly variable. The SN of these observations shows larger variations than in July, reaching values of 7.5.

Spectroscopy within the \emph{Elsie} and \emph{Angkor} events was obtained with the
% GK - I thought the HARPS_N spectrum was in Elsie
HARPS-N fibre spectrograph \citep{harps} attached at the Telescopio Nazionale Galileo (Roque de los Muchachos Observatory).  We covered the spectral range from 383 to 693 nm with a spectral resolution of 115000. The integration time for each spectrum was 15-45 min (see Table \ref{tabla}), reaching a SN around H$_\alpha$ of 30-50. We obtained one spectrum during the egress of \emph{Elsie} and during the minimum and egress of \emph{Angkor}. In order to compare the HARPS-N spectra with a quiescence one, we obtained such observations in d$\sim$ 58056. This spectrum was out of any dip, though it is difficult to identify the quiescence state of the star (see Fig. \ref{phot_curve}). 

\begin{figure}
\includegraphics[width = \columnwidth]{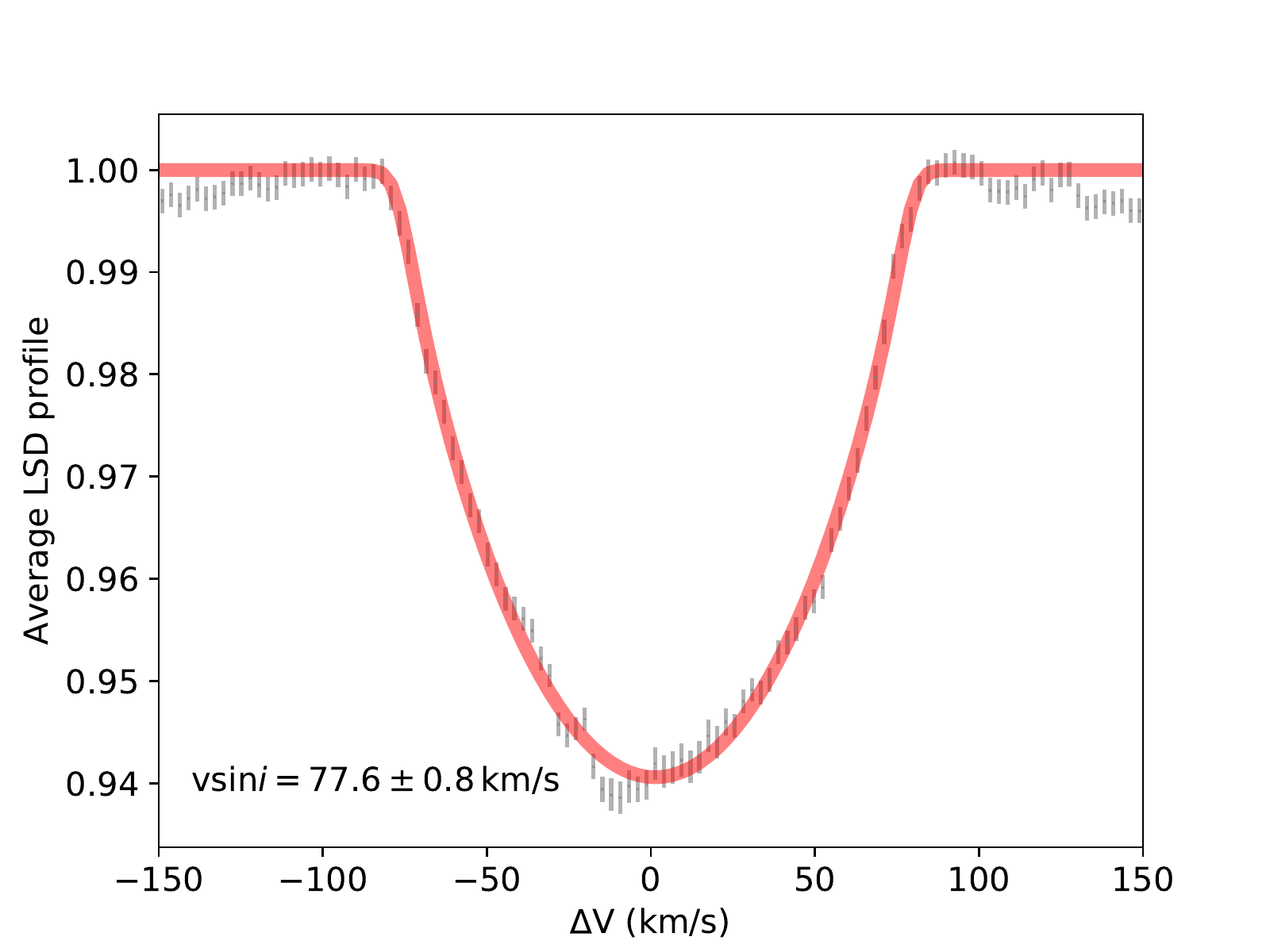}
\caption{Fit, in red, to the average LSD profile from HARPS-N.}
\label{lsd_harps}
\end{figure}

\begin{figure}
\includegraphics[width = \columnwidth]{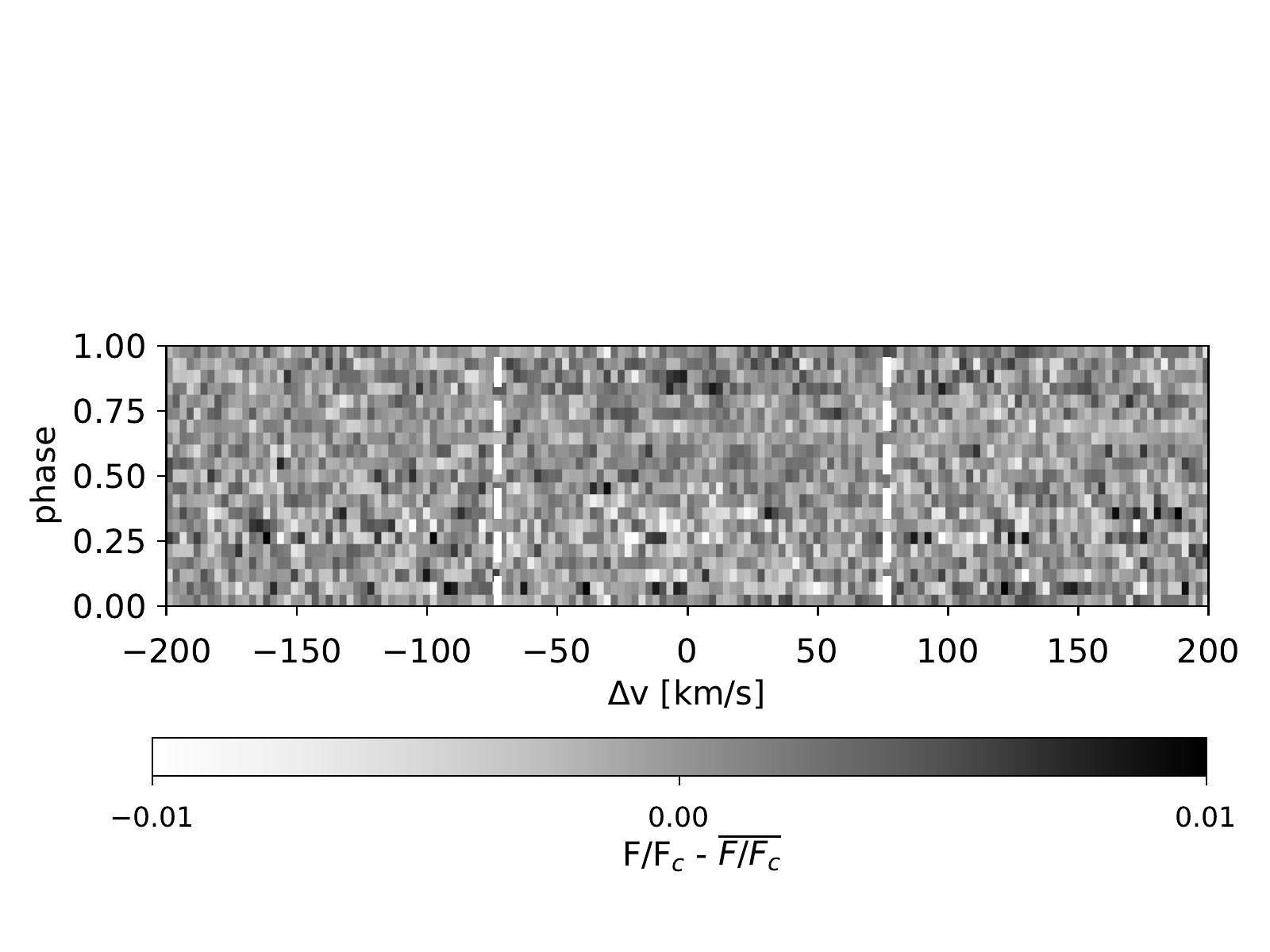}
\caption{Evolution of the LSD residuals with respect to the time averaged profile from HERMES. The phase of the star is normalised to the rotation period, assuming the 0.88 day value. The vertical dashed lines represent the $\pm$v$\sin{i}$ values to delimit the width of the spectral line.}
\label{residuos_lsd}
\end{figure}

We could also obtain 21 medium-high resolution spectra (R = 46000-67000; see Table \ref{tabla} for more details) with the FIES spectrograph \citep{fies} at the Nordic Optical Telescope (NOT) at the Observatorio del Roque de los Muchachos scattered over the period from May to September. The observed spectral range covered from 364 to 911 nm. The integration time per spectra was 17-25 min, though we co-added two spectra to obtain a SN around H$_\alpha$ of about 40. 

The reduction of HERMES, FIES, and HARPS-N data were performed with the standard pipelines provided by the instrument developers. After this automatic reduction, we normalised the non-order-merged spectra by comparing them with a model convolved with a rotational profile appropriate to the star. To overcome the low SN ratio of individual spectra, we build the photospheric average spectral line following the Least Squares Deconvolution (LSD) technique \citep{donati_lsd} following the approach by \cite{andres_lsd}. By so doing, we increase the SN to about 200 for HERMES and FIES data, and to about 300 for HARPS-N.

\begin{figure}
\includegraphics[width = \columnwidth]{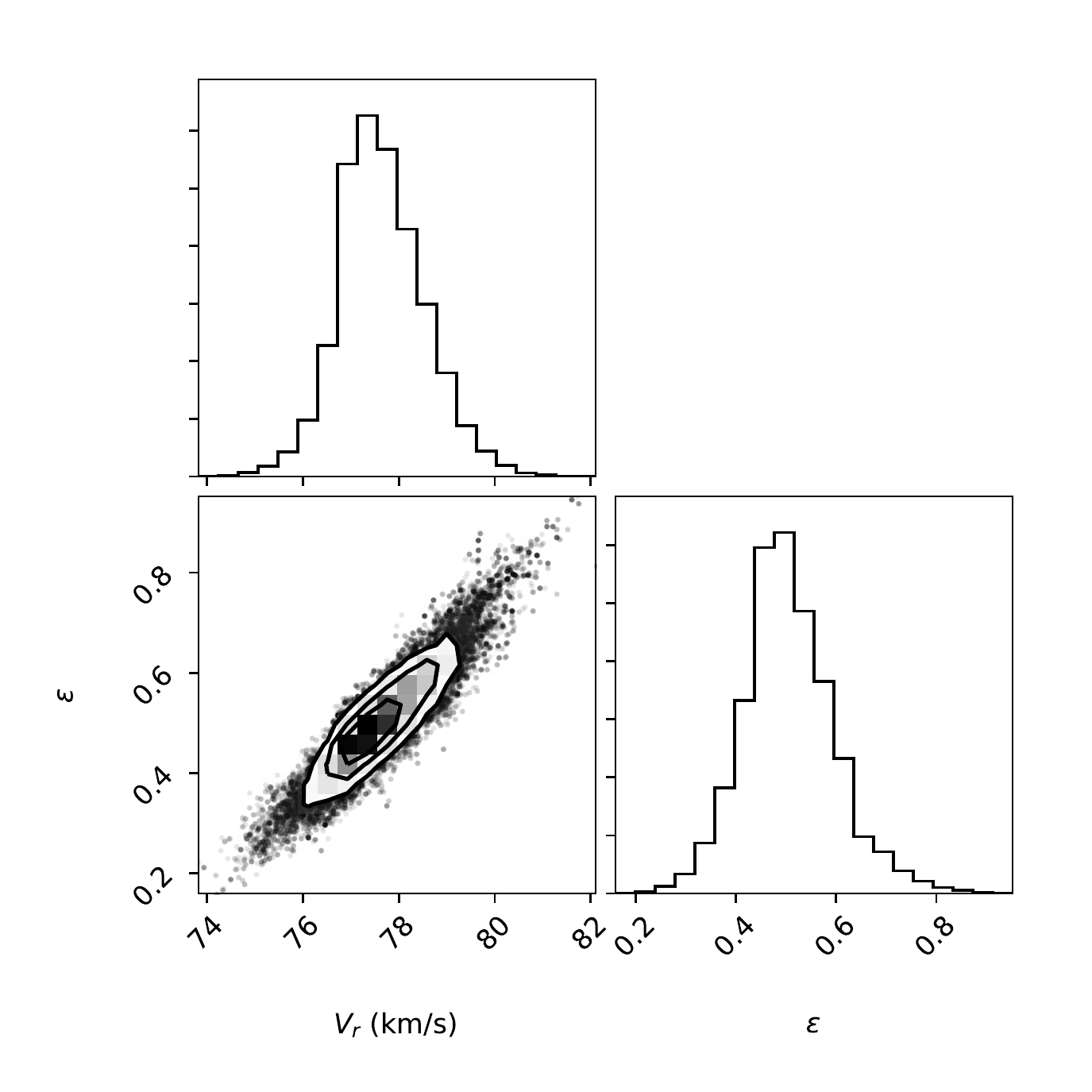}
\caption{Marginalised posterior distributions of the fit to the HARPS-N LSD profile showing the degeneracy between $v_r$ and $\epsilon$.}
\label{posterior_vr}
\end{figure}

\begin{figure*}
\includegraphics[width = \textwidth]{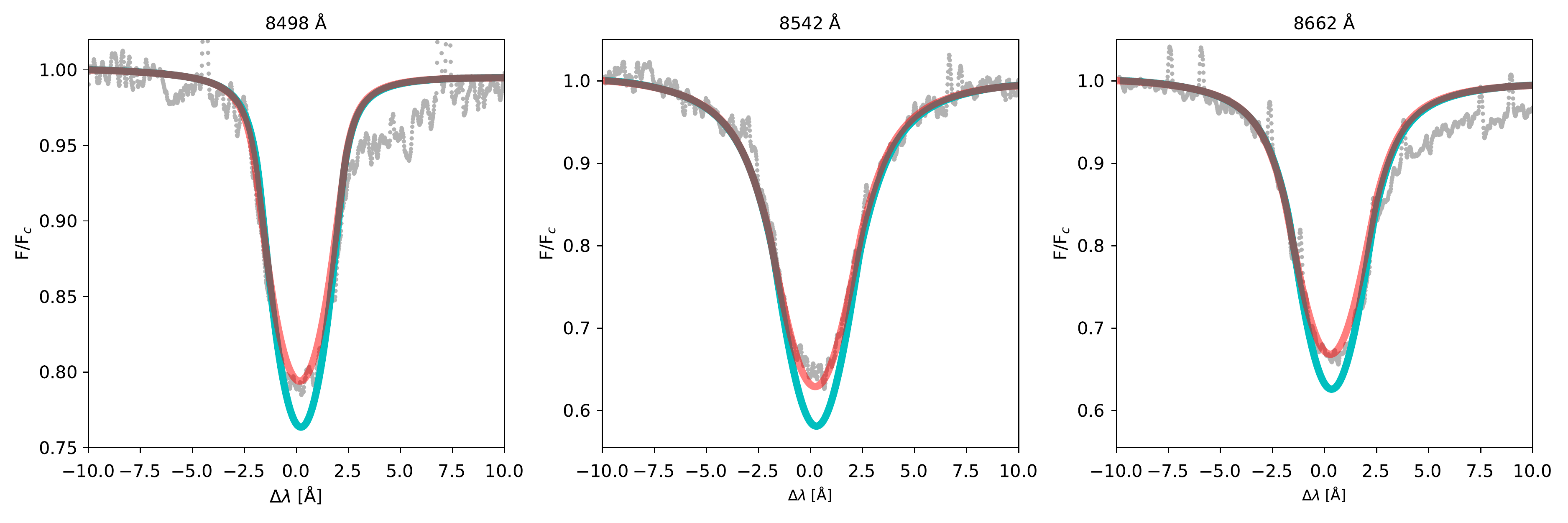}
\caption{Fit (in red) of the Ca\,{\sc ii} infrared triplet lines obtained with the non-LTE inversion code NICOLE. Grey dots represent the observed profiles, while the turquoise line displays the Ca profiles from the radiative equilibrium model that fits the whole HERMES spectrum using the BACCHUS code.}
\label{ca_triplet}
\end{figure*}

\begin{figure}
\includegraphics[width = \columnwidth]{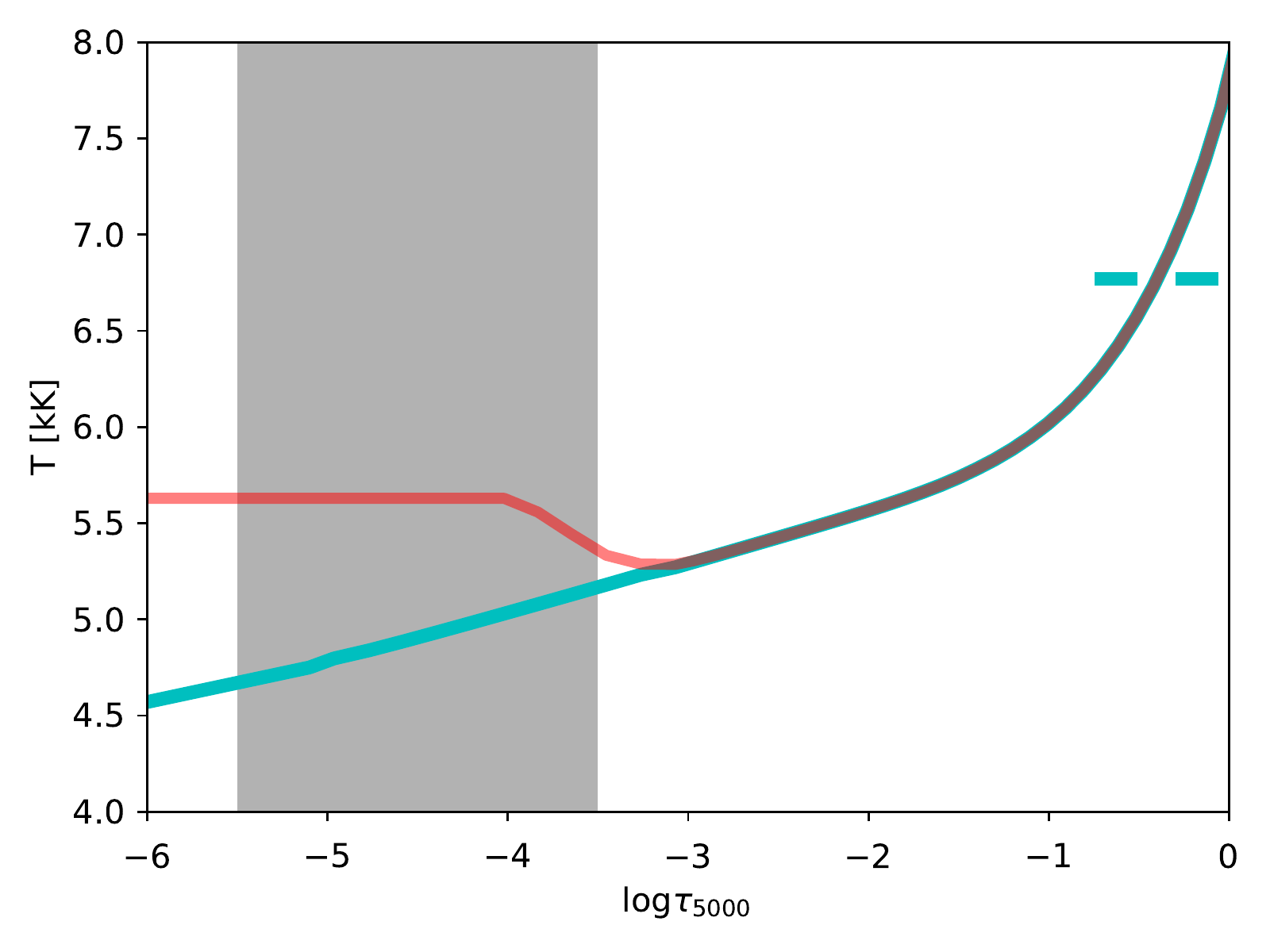}
\caption{Radiative equilibrium model inferred from the whole HERMES spectrum using the BACCHUS code (turquoise) compared to the model atmosphere inferred from the best fit to the Ca\,{\sc ii} triplet lines using the code NICOLE (red line). The horizontal turquoise lines mark the optical depth and temperature of the surface, and the grey area defines, approximately, the formation region of the Ca\,{\sc ii} triplet lines.}
\label{nicole_model}
\end{figure}

\begin{figure}
\includegraphics[width = \columnwidth]{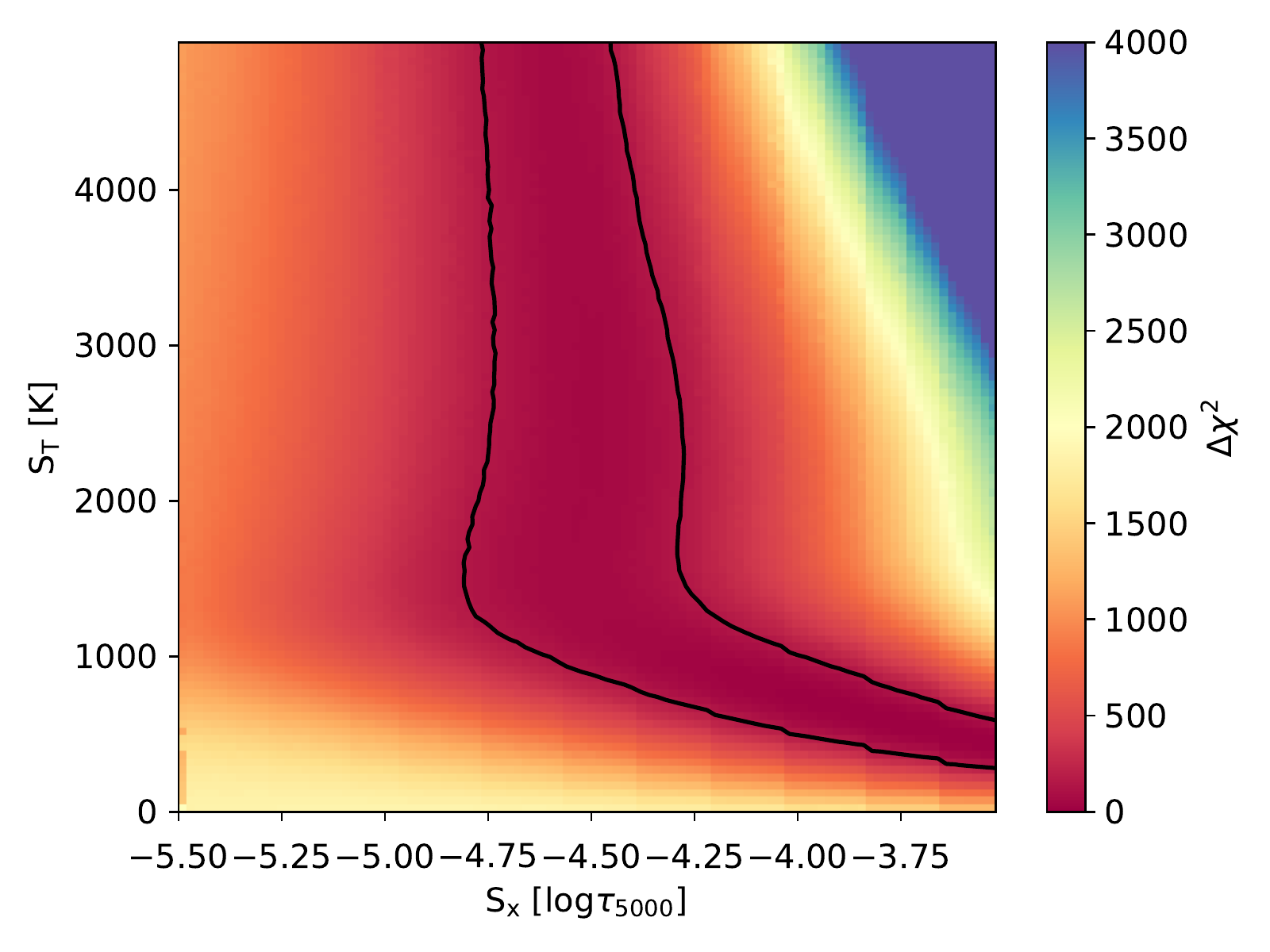}
\caption{Difference between the merit function ($\chi^2$) with respect to its minimum value (the model that better fits the observations). This value ($\Delta\chi^2$) is plotted versus the two free parameters of the model. The black contour delimit the one sigma confidence level.}
\label{matrix}
\end{figure}

\section{Characterisation of the star's atmosphere}

\subsection{Inference of global stellar parameters}
\label{lsdprof}
In order to characterise the stellar parameters of Boyajian's star we build an average spectrum from all the HERMES observations. We choose the HERMES spectra since this instrument covers a wider range of wavelengths than HARPS-N, while still offering comparable spectral resolution. Also averaging over 115 spectra allow us to decrease the noise level significantly. The stellar parameters were recovered with BACCHUS (Brussels Automatic Code for Characterising High accUracy Spectra).  This code relies on the grid of MARCS model atmospheres \citep{marcs}, the model atmosphere thermodynamic structure interpolator developed in \cite{thesis_masseron}, and the radiative transfer code Turbospectrum \citep{turbospectrum,turbospectrum2}. It offers the $T_{\mathrm{eff}}$, $\log{g}$, and metallicity of a given spectrum from fixed values of the projected rotational velocity $v\sin{i}$ and the centre-to-limb-variation coefficients. To obtain these latter parameters, we perform a Bayesian inference over the time averaged LSD from HARPS, as the spectra taken with this instrument have the best spectral resolution. The model proposed to fit the average LSD spectrum is based on \citet{graybook}, and it is the result of the convolution of a Voigt profile with a rotational profile that includes the integrated limb darkening function. This function has the form I$_c$/I$_c^0$=(1-$\epsilon$)+$\epsilon\cdot \cos{\theta}$ and a single free parameter $\epsilon$, the limb-darkening coefficient.

Figure \ref{lsd_harps} shows the average HARPS-N LSD profile and the fit of the maximum a posteriori values for the profile model. We recover $v\sin{i}=77.6\pm0.8\,$km/s and $\epsilon=0.5\pm0.2$ (these are $1\sigma$ confidence intervals). The rotational velocity is slightly lower than in previous works \citep[84.4 $\pm$ 4 km/s][]{BoyajianWTF}, but this parameter is slightly degenerate with the limb-darkening (see the posterior distributions in Fig. \ref{posterior_vr}), so depending on how this is modeled, the discrepancies can be brought into concordance. We derive a value of $8\pm5\,$km/s for the FWHM of the Gaussian contribution to the Voigt profile. The FWHM of the Lorentzian is poorly constrained at this SN, and we just recover the prior. However, though not constrained by the observations, their uncertainty is not translated to the rest of parameters that are uncorrelated.

Although line-of-sight (LOS) velocity determinations will be discussed later (Sec. 5), we use all the available spectra to derive an average value (by optimally weighting with each individual uncertainty) of $v_\mathrm{los}=4.21\pm0.06\,$km/s. This average includes the spectra from \citet{BoyajianWTF}.

With this value for $v\sin{i}$, we use the average HERMES spectrum and BACCHUS to derive the physical parameters for the star, i.e., $T_{\mathrm{eff}}=6750\pm 180\,$K and $\log{g}=4.0\pm 0.3$.
These values are compatible with those obtained in \cite{BoyajianWTF}. With the photometry summarised there and the new parallax estimate from Gaia DR2 \citep[$p=451\pm5\,$pc,][]{gaiadr2} we can derive the bolometric magnitude for Boyajian's star, using the bolometric corrections from \citet{bolo}. We obtain $M_\mathrm{B}=3.15\pm0.11$. This error estimate contains both the error in $T_\mathrm{eff}$, in $E(B-V)$, distance and apparent magnitude. This value implies that the radius for the star is $R=(1.53\pm0.34)R_\odot$, and from this and assuming that the modulation in the light curve of 0.88 day corresponds to the rotation of the star, and using our estimate for $v\sin{i}$ we arrive at a viewing angle of $i=62^\circ\pm26^\circ$ \citep[in agreement with the 68$^\circ \pm$29$^\circ$ value found by][]{BoyajianWTF}.

\subsection{Surface magnetic activity}

The \emph{Kepler} light curve of Boyajian's star shows a clear modulation of the flux with a period of 0.88 day and an amplitude of $\sim$ 0.1\% that is present throughout the time series (1500 days). The star is located at the edge of the instability strip of $\gamma$-Doradus-like pulsators, and the period is compatible with the ones observed in it \citep[e.g.][]{uytterhoeven11}. However, the fact that the amplitude of this oscillatory pattern is highly variable in time, and that this period is compatible with the rotation inferred from the broadening of spectral lines, \cite{BoyajianWTF} claim that this period is likely due to stellar activity, i.e., brightness inhomogeneities in the surface due to starspots or bright plage areas. 

We check this hypothesis by studying the variability of the photospheric lines and its relation to the rotational period (assuming it is 0.88 day).  Since in our spectra individual spectral lines have low SN, we use the LSD photospheric profile to search for variability due to inhomogeneities in the stellar surface. Since we are not seeing the star from one of its poles (in fact, the viewing angle is quite large), surface dark spots leave a pattern in the spectral line similar to an emission that moves across the line profile from the blue wing to the red one. For an equatorial spot, the pattern moves from $-v\sin{i}$ to $+v\sin{i}$ during half the rotation of the star. The smaller the velocity range spanned, the higher the latitude of the spot. Fig. \ref{residuos_lsd} shows the evolution of the residuals of the LSD profiles during all phases of the star (normalised to the rotation period) obtained in the July campaign -- hence with higher SN and not affected by the dipping event \emph{Elsie}. The residuals are computed as the difference of the LSD profiles and the time averaged LSD. To build the Figure, we have averaged all the spectra within a phase bin of 0.05. There is no spot-like pattern in the evolution of the LSD profiles. 

The lack of such signature means that either the surface is homogeneous or that it has small spots that are undetected at our noise level. To support these results, the radial velocities of the LSD profile (see Sect. 5) do not show any modulation with the 0.88 day assumed rotation period. To estimate an upper limit to the size of the spots that can be detected given our noise level, we have carried out some numerical simulations. We synthesise a spectral line from a rotating star (with Boyajian's star rotation values) with local Gaussian profiles. The local profiles are either spot profiles or quiet profiles, the difference being the parameters of the Gaussian. We do not take into account any centre to limb variation. Assuming then solar values for a spot, i.e., a continuum contrast of 0.1 with the quiet photosphere, and a depth of the line of 0.6 the spot continuum (i.e. deeper than the quiet photospheric lines), we can reproduce the observed 0.1\% variation of the light curve with a spot size of 0.03 rad. This means a spot that covers 0.02\% of the stellar surface. In the Sun, a typical sunspot represents 0.002\% of the solar surface. Larger spots would be incompatible with the \emph{Kepler} photometry. This simulated spot would produce modifications in the global profile (which is similar to the LSD profile of Boyajian's star) of $\sim 3\times 10^{-4}$ F$_c$. Therefore, if spots exist in the surface of Boyajian's star, they are undetectable in high resolution spectroscopy at our sensitivity ($\sim 10^{-3}$ F$_c$). Also, note that we have performed the tomography by combining spectra taken during 6 nights. This implies almost 7 rotations of the star. If spots have shorter lifetimes, it would make them even more difficult to detect.

\begin{figure}
\includegraphics[width = \columnwidth]{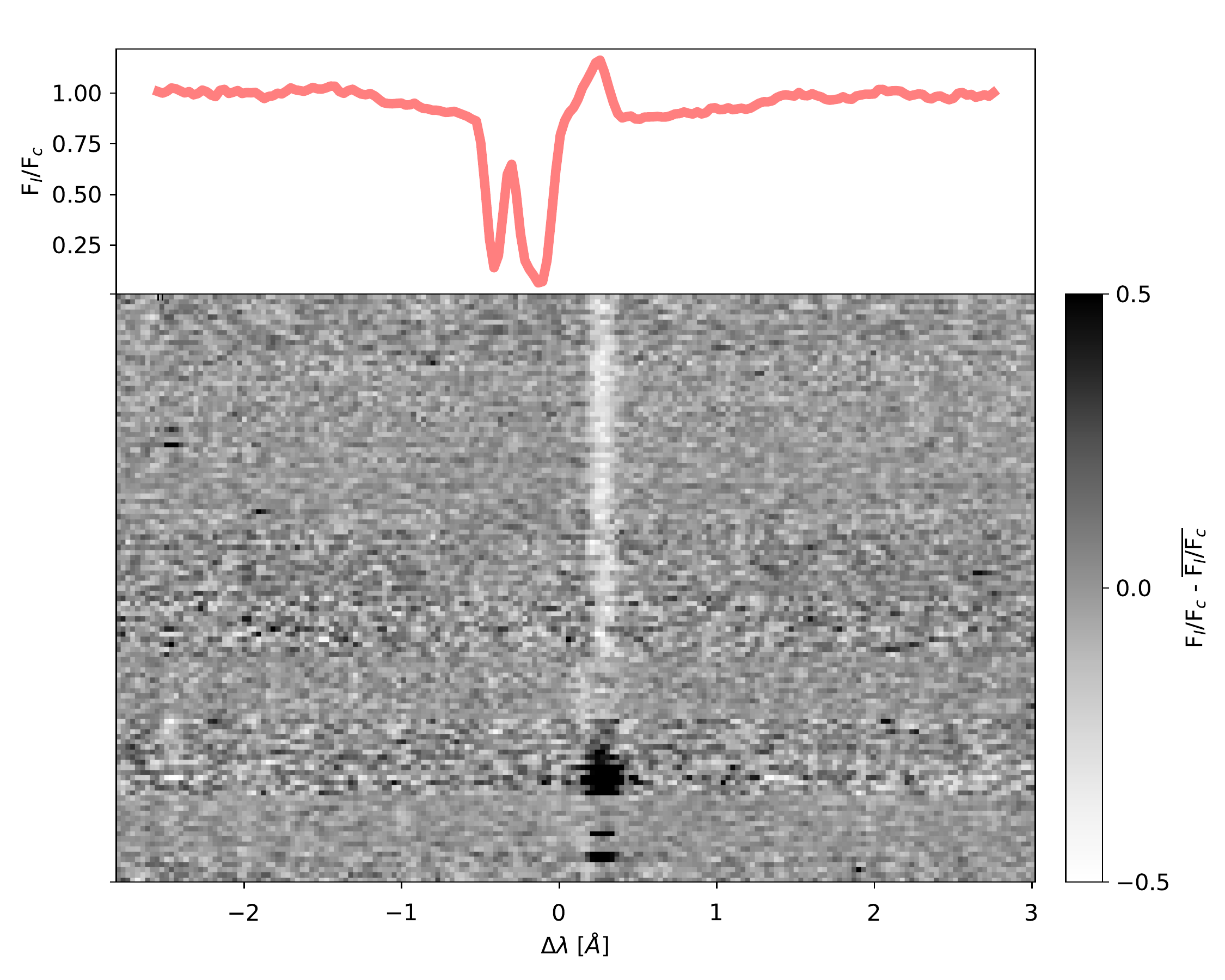}
\includegraphics[width = \columnwidth]{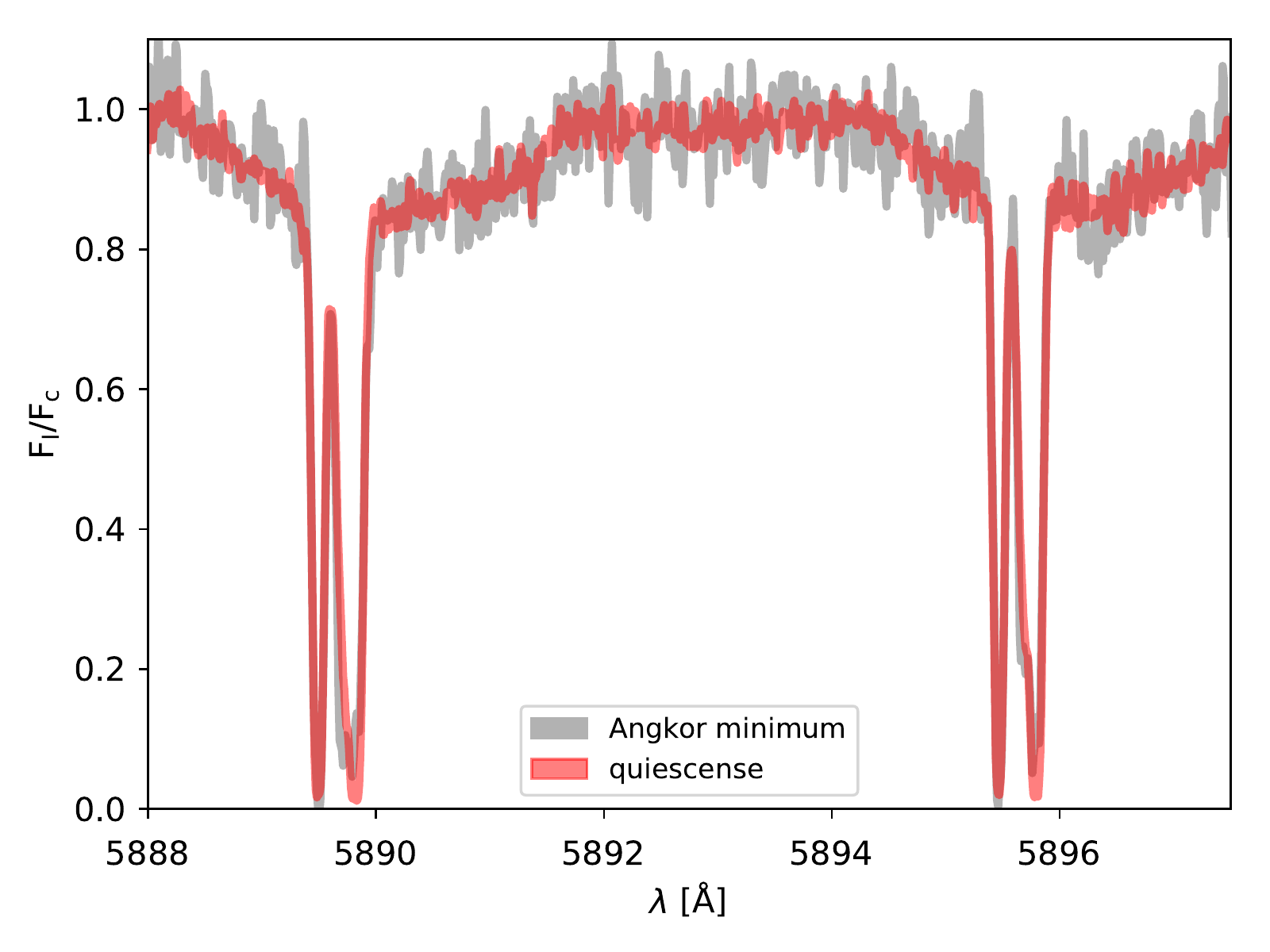}
\includegraphics[width = \columnwidth]{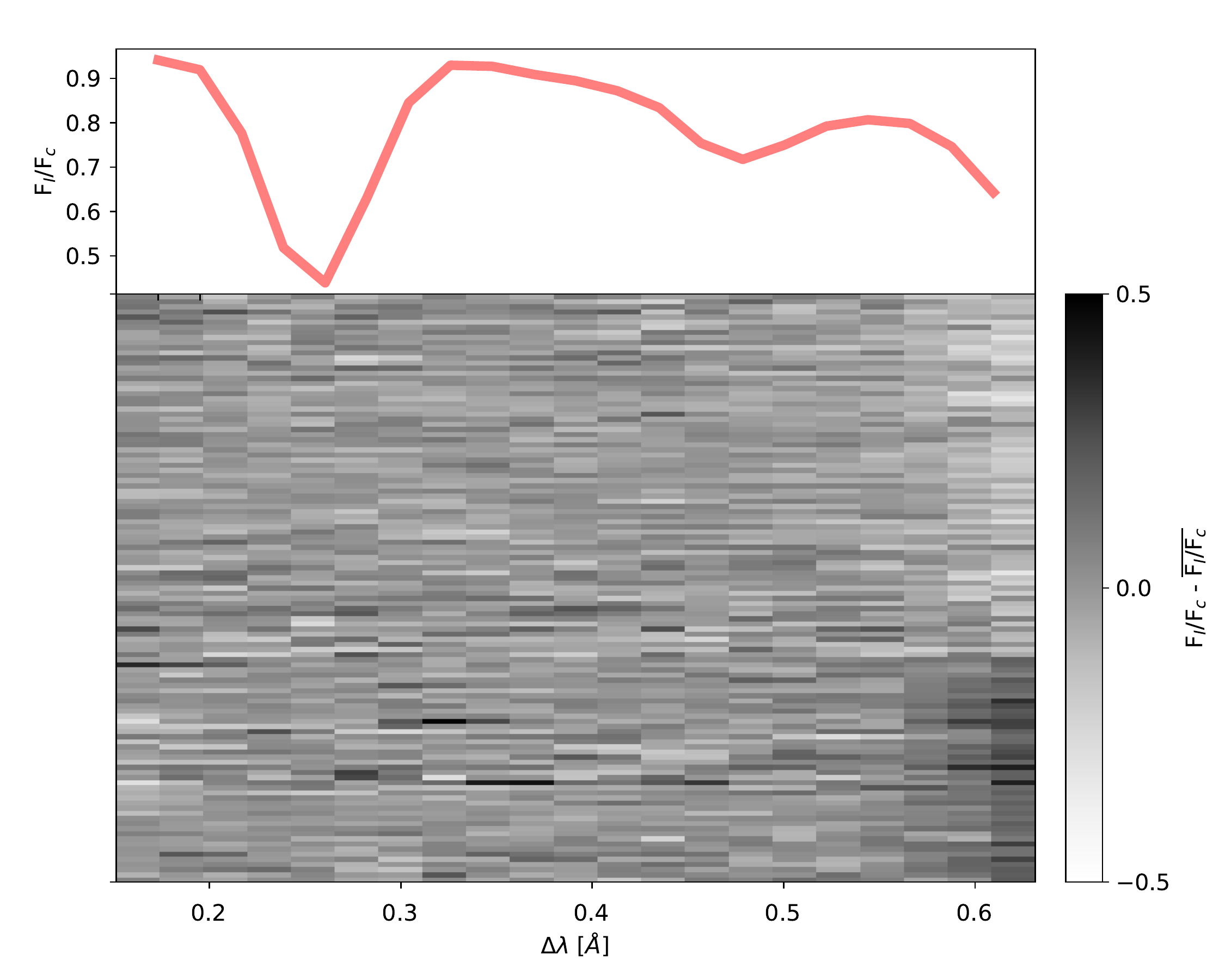}
\caption{Top, bottom panels: time evolution (vertical axis) of the interstellar Na D1 (Mg I) lines and time average spectra (red lines) from HERMES observations. Note that all the observed spectra have been plotted together and the vertical axis has not a constant time interval. Middle panel: Comparison of the Na D1 and D2 lines observed with HARPS-N in the minimum of the \emph{Angkor} dip and out of the dip (quiescence).}
\label{im}
\end{figure}

\begin{figure}
\includegraphics[width = \columnwidth]{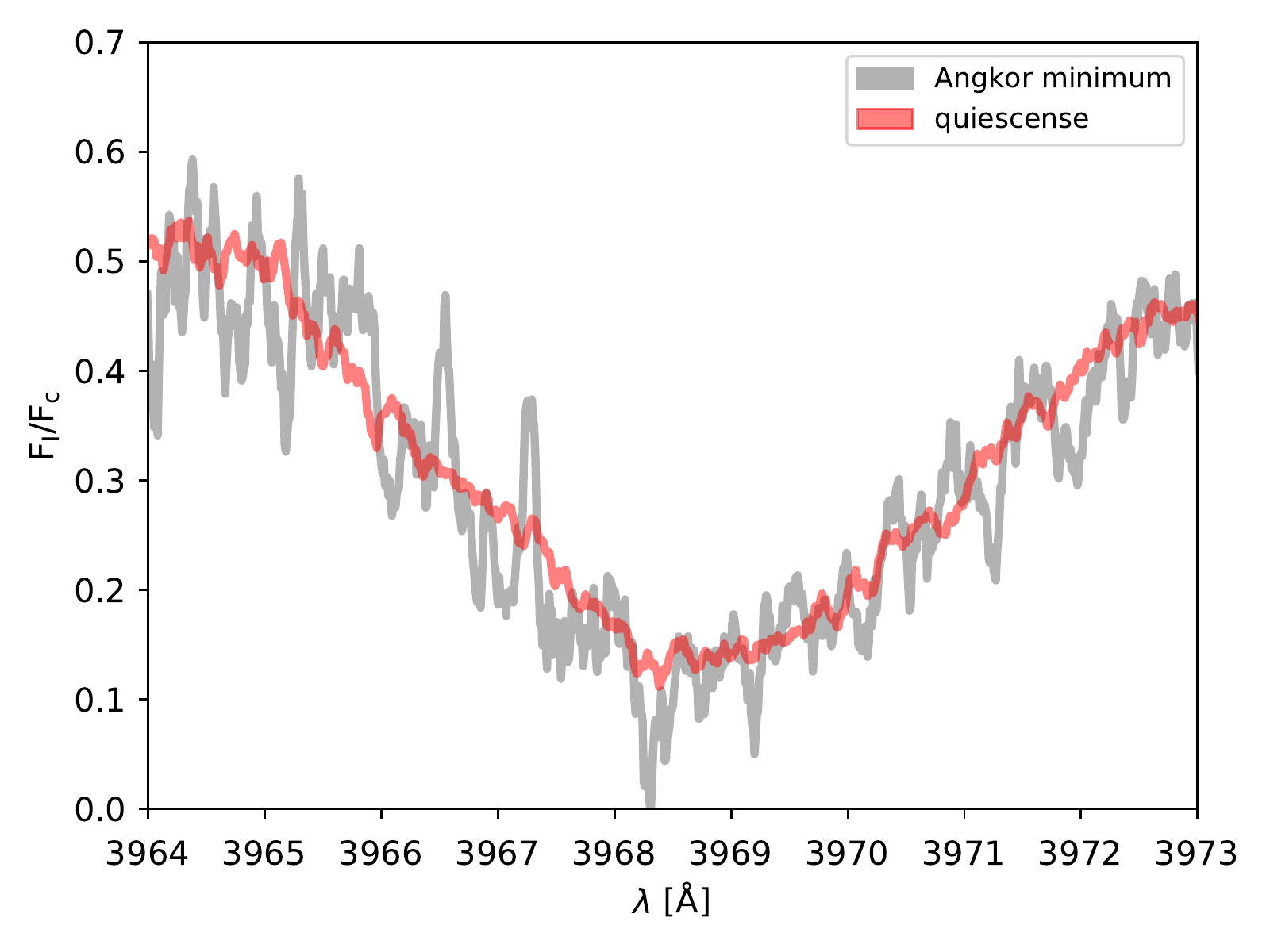}
\caption{Comparison of the Ca\,{\sc ii} H line observed with HARPS-N in the minimum of the \emph{Angkor} dip and out of the dip (quiescence). Since the noise level at these blue wavelengths is high, we have used a running average of 10 wavelength points to display the profiles.}
\label{cahk}
\end{figure}

\subsection{Chromospheric activity}

Other signs of magnetic activity can be found in some chromospheric lines, such as H$_\alpha$ or H$_\beta$, the core of the Na D1 and D2 lines, the He\,{\sc i} D3 line, and the Ca\,{\sc ii} H, K, and infrared triplet lines. In our spectra, the cores of the Balmer lines are contaminated with Earthglow and very likely by some nebular emission. Therefore, they are not as reliable to diagnose the possible magnetic activity of the star. We have found no clear evidence of rotationally-modulated activity in none of the other chromospheric lines. 

However, we notice that the model that better fits the stellar spectrum as a whole is unable to reproduce the cores of the Ca\,{\sc ii} infrared triplet lines at 8498, 8542 and 8662 \AA, which appear raised with respect to this model, as it predicts deeper cores. Fig \ref{ca_triplet} shows a comparison for the \ion{Ca}{II} triplet between the average profile of the HERMES data from our July campaign (grey dots) to a synthetic spectrum obtained from the best fit model to the full spectrum with BACCHUS (turquoise; $T_{\mathrm{eff}}$=6750~K, $\log(g)=4$ and solar metallicity). To take into account the non-LTE effects (and to be consistent with the inversion that will be carried out later) we have performed the synthesis of the Ca\,{\sc ii} lines with the non-LTE code NICOLE \citep{nicole1,nicole2}. Figure \ref{ca_triplet} shows that the wings of the three Ca lines are accurately described by the model but the observed line core intensities are higher in all three lines, even after considering the departure of LTE. The wings of the lines trace the photosphere of the star, where temperature decreases due to radiative losses. Closer to the core, the lines trace higher (chromospheric) layers, where the temperature continues decreasing in the radiative equilibrium model but it is known to increase in some stellar atmospheres. 

We investigated the temperature stratification of Boyajian's star using NICOLE in inversion mode to fit the Ca\,{\sc ii} infrared triplet lines. Generally speaking, all of the chromospheric information in these lines is encoded within the relatively narrow central core which, for this star, would have a width of $\sim$5~km~s$^{-1}$ in absence of rotation. Unfortunately, much of the information is washed out in the convolution with the much broader rotational kernel, which severely limits the amount of detail that we may hope to infer from the observations. For this reason, instead of aiming for a reconstruction of the full stratification (the typical operating mode of NICOLE), we now reduce the stratification to a photosphere fixed by the model and a chromosphere that is defined by a sharp temperature increase. In this scenario we only have two free parameters that control the chromospheric temperature, hence the cores of the lines: the height at which the step occurs (S$_\mathrm{x}$) and its amplitude (S$_\mathrm{T}$). The resulting best fit to the spectral lines is shown in red in Fig.~\ref{ca_triplet}, while the inferred atmospheric stratification corresponding to the best fit is shown in Fig~\ref{nicole_model} (red line). We conclude that the chromosphere of Boyajian's star has a higher temperature than would be expected in radiative equilibrium (turquoise). The best fit corresponds to S$_\mathrm{x}=-4.6$ (in continuum optical depth a 5000~\AA ) and S$_\mathrm{T}=1250$ K. However, we found that some degree of degeneracy exists between S$_\mathrm{x}$ and S$_\mathrm{T}$ even at a 1-sigma level (see Fig. \ref{matrix}). The figure reveals two families of solutions. The first has the temperature step between $\log(\tau_{5000}$=-3.5 and -4.5 and a temperature increase between 500 and 1000~K above the reference (non-active) model. The second family is seen as a vertical stripe in the figure. If the temperature increase occurs higher in the atmosphere than $\log(\tau_{5000})$=-4.75, then the observations are compatible with any temperature increase above 1000~K. Synthetic profiles with no rotation show variations in the structure of the emission reversals formed in the Ca infrared triplet lines but the large rotational broadening dilutes this information making it impossible to distinguish between models with vastly different temperatures. Despite this degeneracy, the fact that the outer atmosphere of Boyajian's star is hot is well constrained by the observations.

Stellar activity can be understood as the presence of hot outer atmospheres, i.e.,  the lack of steady-state balance of input radiative and convective energy from the stellar interior and radiation to space \citep{linsky2017}. The emission in the infrared triplet is an indicator of stellar activity \citep{andretta05} that is correlated with the widely-used R$_{HK}$ index -- the emission of the core of the Ca\,{\sc ii} H and K lines \citep{busa07}. We check that the Ca\,{\sc ii} H and K core emission in our data is compatible with the one observed in the Ca\,{\sc ii} infrared triplet. To do so, we synthesise the Ca\,{\sc ii} H and K lines using the model obtained with the fit to the Ca\,{\sc ii} infrared triplet. The synthetic profiles agree with the observed Ca\,{\sc ii} H and K lines within the noise level, which is indeed much higher than in the infrared triplet.

In this way we can then assert that Boyajian's star has some degree of activity, i.e., it harbours a hot chromosphere. Interestingly, this level of activity seems to be intrinsic to the star since it is constant along all our observations and it is clearly not associated to the flux dimmings. 

Most if not all F to M dwarfs, giants, and supergiants are active at some level. But the heating source of the outer stellar atmospheres is still a matter of debate and even a long-standing problem in Solar Physics. Raised cores of the Ca\,{\sc ii} infrared lines, as well as emission in the cores of the Ca\,{\sc ii} H and K lines, are commonly observed in magnetised areas of the solar chromosphere \cite[e.g. the review by][]{review_jaime}, because the vertical fields collimate acoustic and magneto-acoustic oscillations from the photosphere that dissipate in the chromosphere via shocks. Global acoustic oscillations of the Sun traveling across the whole atmosphere can cause a global heating of the outer layers \citep[e.g.]{nazaret}. Also magnetic reconnection and dissipation can be a cause of the global chromospheric heating \citep{javier,marian_loops}. What could be causing the heating of Boyajian's star chromosphere is beyond the scope of this paper but it is likely an imprint of global magnetic activity (since we do not observe a modulation of the core intensity of the Ca\,{\sc ii} triplet with the 0.88 day period), wave propagation, or both. 

\section{The hypotheses of interstellar clouds and falling evaporating bodies}

If some sub-AU structuring of the interstellar medium were responsible for the obscuration of Boyajian's star \citep{Wright2016}, we would expect to detect variations in the depth, width or radial velocity of some spectral lines coming from interstellar clouds. We focus on the Na D1 and D2 lines and on the Mg\,{\sc i} line at 766.4 nm. 

As shown in \cite{BoyajianWTF} and \cite{Wright2016}, the absorption of the Na D1 and D2 lines by the gas counterpart of the interstellar medium has three components, one of them saturated. Our HERMES/HARPS-N/FIES data show no variation of this scenario (see Fig. \ref{im}). Top panel of Fig. \ref{im} shows the time evolution of the residual of the Na D1 line with respect to the time averaged spectrum from HERMES observations. There is no clear variation of the line profiles of the interstellar medium Na lines. Similarly, we do not detect any variation in the radial velocities (see Sect. \ref{rvs}) of these Na lines. The middle panel of Fig. \ref{im} shows the comparison of the HARPS-N spectra of the Na D1 and D2 lines within the minimum of the \emph{Angkor} dip (the deepest one in the time series) and out of the dip. As can be seen, both spectra are similar within the noise level in and out of dip. To further check the apparent non-variation of the interstellar medium lines, we also study the Mg\,{\sc i} line at 7664 \AA.
Bottom panel of Figure \ref{im} shows the time evolution of the residuals of the Mg\,{\sc i} line. As in the case of the Na lines, there is no variation (also no variations of the radial velocities). Hence, we can safely assert that the variation at sub-AU scales of the interstellar gas is not producing the dimming events of the star. Though, we can not rule out the role of the dusty interstellar medium on obscuring the star since its evolution can be decoupled from the gas \citep{jason}.

Falling evaporating bodies (as well as exocometary tails) can be detected in ionised species, in particular, they have been reported in the Ca\,{\sc ii} H and K lines \citep[e.g.][and references therein]{kiefer2014}. We look for the same signatures in the ionised Ca lines, including also the infrared triplet, but we find no evidence. Figure \ref{cahk} shows the comparison of the Ca\,{\sc ii} line recorded with HARPS-N during the minimum of the \emph{Angkor} event and out of the dip. At least at the noise level of our observations, there are no traces of falling evaporating bodies as seen in \cite{kiefer2014}.

\section{Radial velocities. The hypothesis of a clumpy, dusty environment}
\label{rvs}

We derive the LOS velocity of each single-epoch LSD profile \citep[not only HERMES spectra, but also HARPS and FIES, including those from][]{BoyajianWTF} as described in Sec. \ref{lsdprof}, but also leaving v$_\mathrm{los}$ as a free parameter. For this computation, all spectra have been corrected by the barycentric velocity. We choose an uninformative uniform prior for $v_\mathrm{los}$, as these are amenable for location parameters such as this one. The $1\sigma$ confidence intervals for v$_\mathrm{los}$ depend strongly on the SN, but typically for a 30 min spectrum taken with HERMES these have $\sigma=0.8\,$km/s. The other inferred parameters are coherent with what has been discussed in Sect. \ref{lsdprof}, although much more poorly constrained due to the lower SN. For HARPS the typical uncertainty is of $0.3\,$km/s. In the case of FIES, for the spectra taken during 2017 the uncertainty is at $0.7\,$km/s, while for the longer exposure spectra from \citet{BoyajianWTF} these drop to $0.3\,$km/s. Using LSD profile fitting we recover an average value of $v_\mathrm{los}=4.1 \pm 0.2\,$km/s, for the spectra from \citet{BoyajianWTF}. This value is in concordance with their previous determination based in cross-correlation techniques. 

\begin{figure}
\includegraphics[width = \columnwidth]{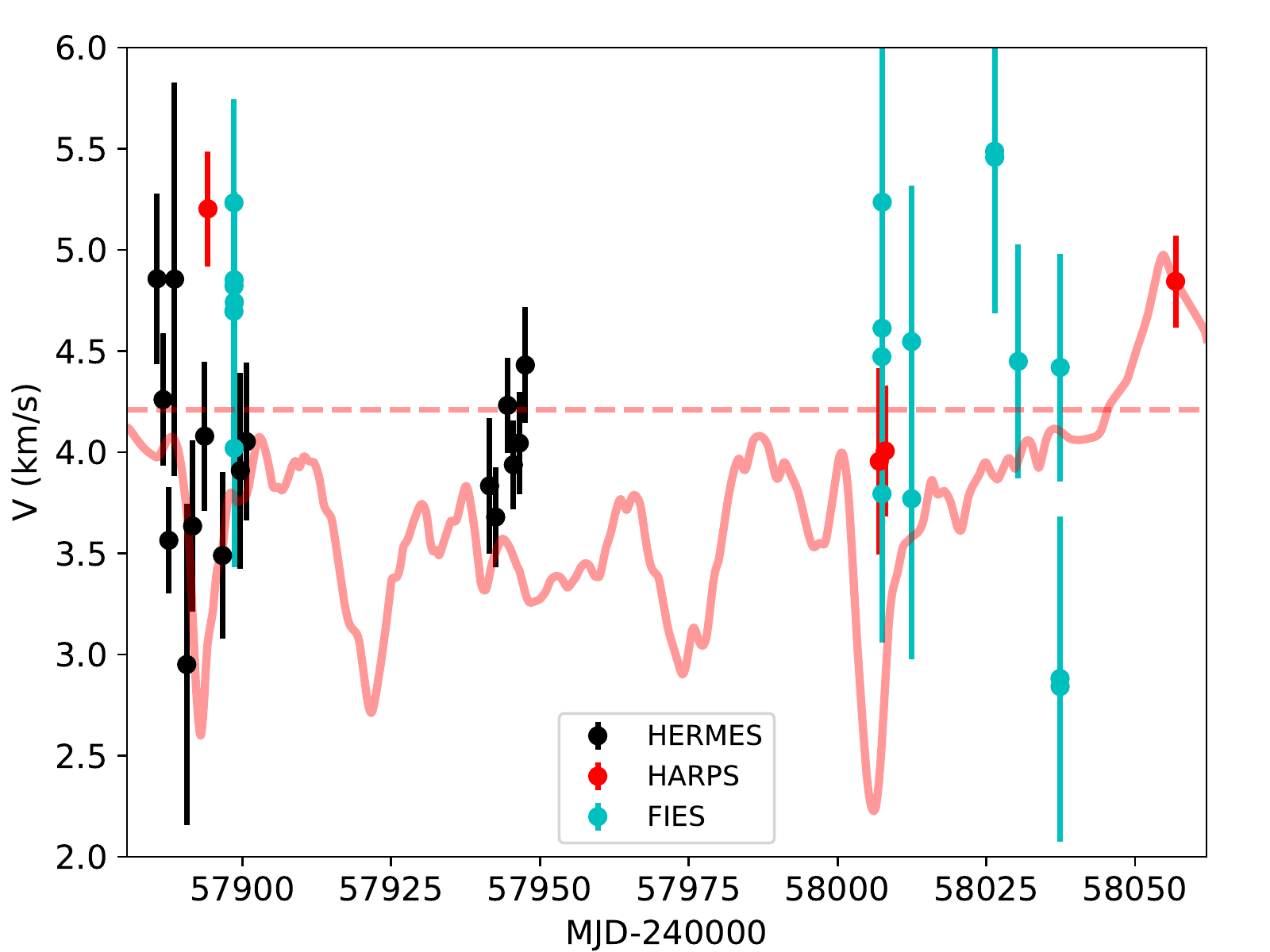}
\caption{Variation of $v_\mathrm{los}$ as measured with HERMES/FIES/HARPS-N using different colours. The rescaled light curve from Fig. \ref{phot_curve} is shown for reference. The dashed horizontal line marks the velocity at a quiescent state (see text).}
\label{vel_rad_all}
\end{figure}

\begin{figure}
\includegraphics[width = \columnwidth]{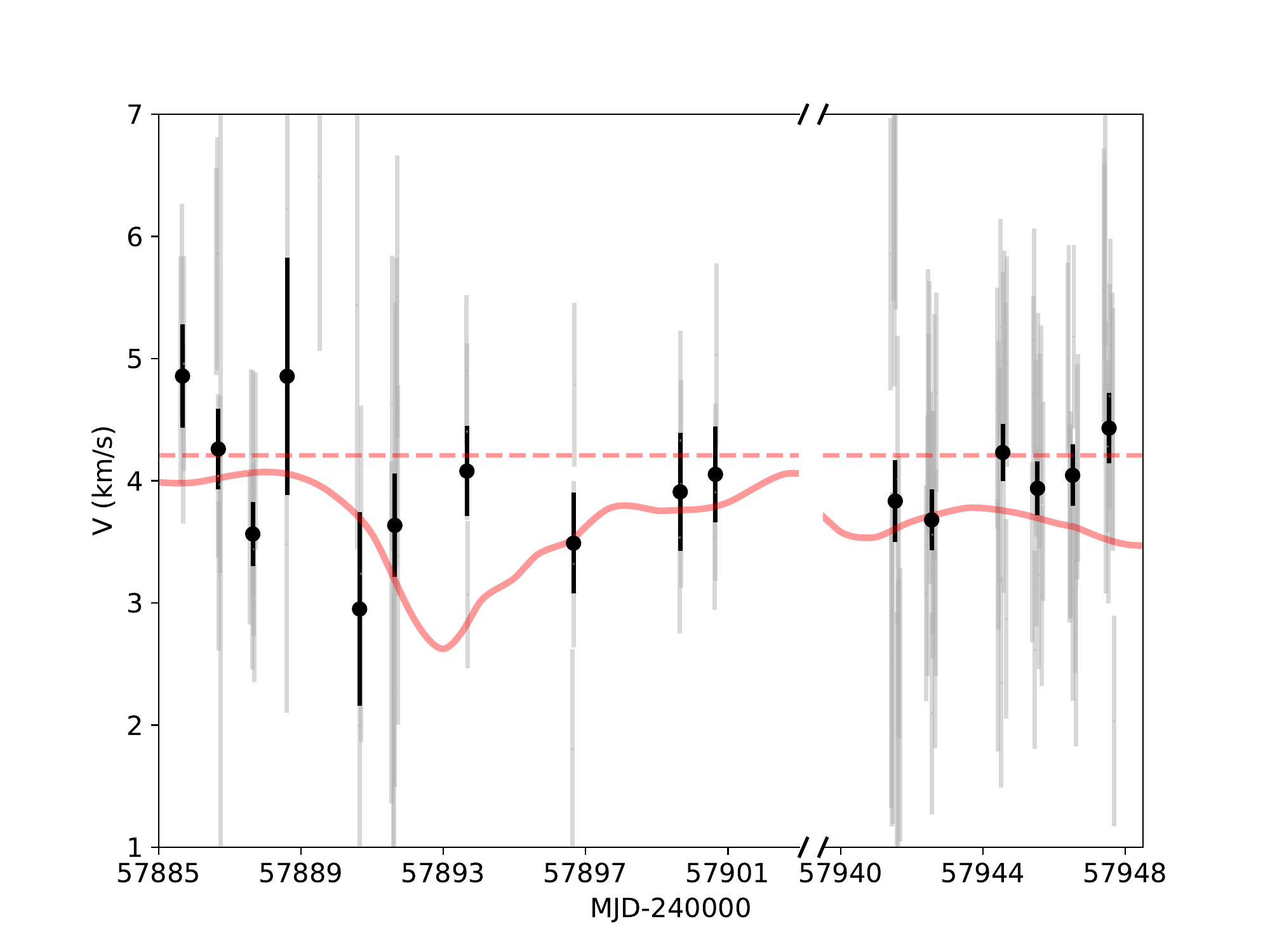}
\caption{Same as Fig. \ref{vel_rad_all} but only for HERMES campaigns. The vertical grey bars mark the values derived from individual spectra while the black dots are averaged daily values. }
\label{vel_rad_mer}
\end{figure}

\begin{figure}
\includegraphics[width = \columnwidth]{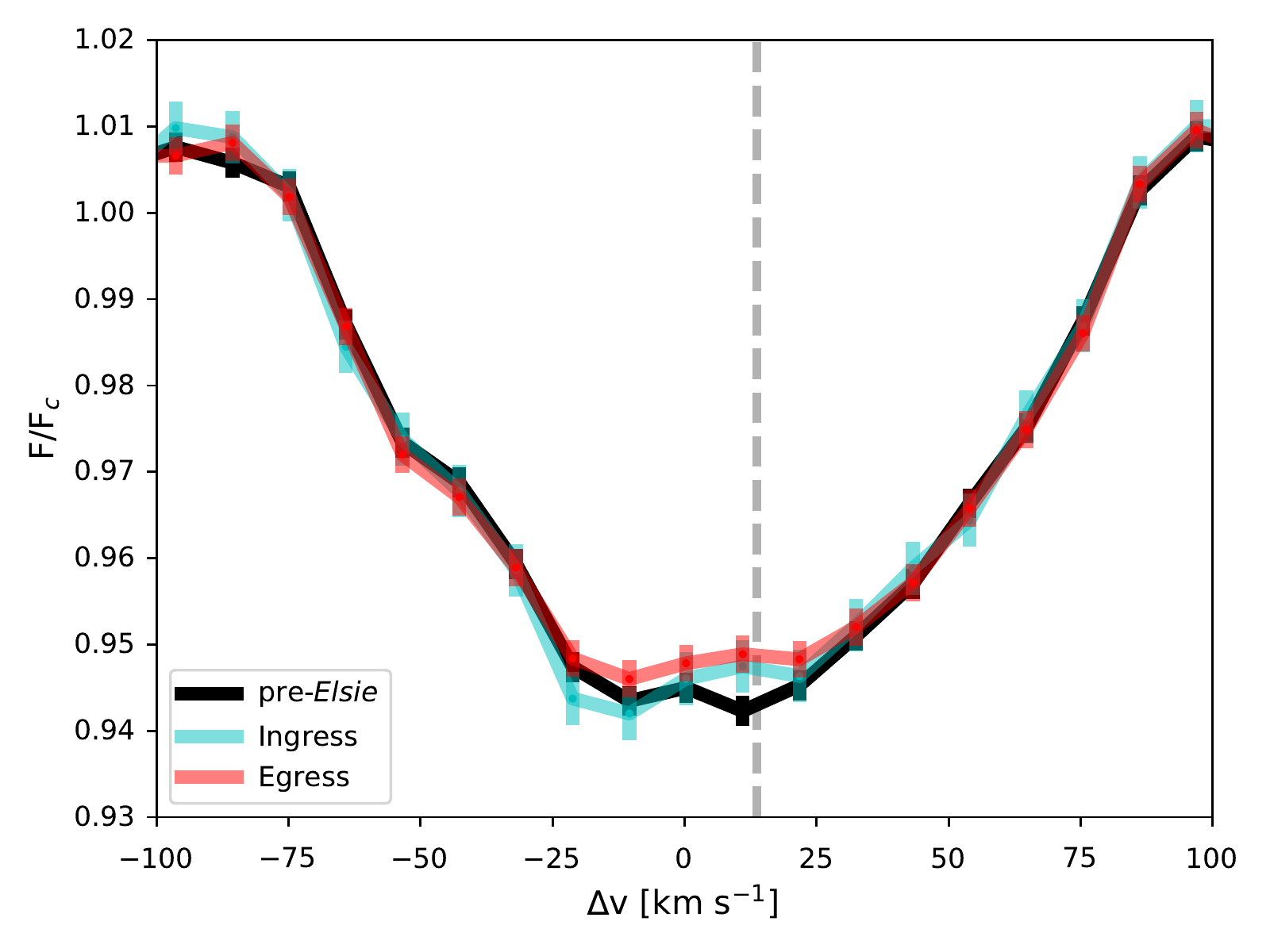}
\caption{LSD HERMES profiles during the \emph{Elsie} event. The vertical dashed line represents the velocity of the Sun relative to Earth.}
\label{lsd_elsie}
\end{figure}

As the cadence of our HERMES spectra is high and the SN is low, in order to facilitate comparison with other datasets, we produce nightly averaged values. As the flux variations last for several days, we can still potentially map the effects of whatever mechanism lies behind this behaviour into the measured v$_\mathrm{los}$. Figure \ref{vel_rad_all} displays the LOS velocities for all the spectra. The visible light curve of the star published in \cite{BoyajianTEAM2018} is superimposed to give context to the velocity measurements. The HARPS-N and FIES spectra are scattered in time and it is difficult to extract any strong conclusion. The most suitable data to obtain reliable conclusions are the HERMES data since they fully sample the \emph{Elsie} event and a flat but lower luminosity interval after \emph{Celeste}. Figure \ref{vel_rad_mer} shows a zoom on region of Fig. \ref{vel_rad_all} but containing only the HERMES data. If we consider the weighted average velocity (4.21 km/s, marked as a horizontal dashed line) as being the LOS velocity in quiescence, we see that the night averages velocities prior to the \emph{Elsie} event are consistent with this value within the error bars. Interestingly, during \emph{Elsie}, the LOS velocities have lower values at a 1 to 2 sigma level. These velocities are not a shift of the spectral line but the effect of a deformation in the line profile. Figure \ref{lsd_elsie} shows the averaged LSD HERMES profile pre-\emph{Elsie} (black) as compared to the profiles within the dipping event (turquoise for the ingress and red for the egress). During \emph{Elsie} the deformation of the line is similar, consisting on an emission feature in the red-shifted wavelengths near the core. Though an accurate modeling is restrained by the SN, this behaviour is suggestive of an optically thick body absorbing/occulting the surface of the star at such red-shifted velocities. 

When a star is eclipsed by an optically thick body, the spectral lines present less absorption at the wavelengths coinciding with the rotational projected velocities of the portion of the occulted surface. When computing radial velocities (assuming that the stellar profile is unperturbed) the profile seems to be shifted towards the wavelengths of the non-occulted surface. This effect is known as Rossiter-McLaughlin \citep{rossiter,mclaughlin}. We claim detection of this effect during \emph{Elsie} with moderate statistical significance, at a 1-2 sigma level. Since the velocities during \emph{Elsie} are all lower than the pre-Elsie values (blue-shifted), the occulted surface during the eclipse is always red-shifted. This implies that the occulting body has a very inclined orbit with respect to the stellar equator. Also, very likely, the impact parameter is high, i.e., the occultation is taking place close to the stellar limb. Since it has been shown that the dimming events are chromatic and likely due mostly to optically thin dust \citep{BoyajianTEAM2018}, we claim that the dusty environment around the Boyajian's star has clumps of optically thick material. 

The radial velocities after the \emph{Celeste} event show a rise to the quiescent value. The photometry does not show a clear dip but a more or less constant flux at lower levels than the pre-\emph{Elsie} values. 

\section{Discussion and conclusions}

We have presented a detailed study consisting of high resolution spectroscopy during obscurations of Boyajian's star. Being  a low magnitude star, performing high resolution spectroscopy with high SN ratio poses a significant challenge. During our campaigns the weather conditions have been far from optimal, 
not enabling us to obtain perfect quality data. However, even at our moderate SN ratio, we have been able to characterise the stellar atmosphere and to help constrain the models proposed in the literature that may be causing the dimmings of Boyajian's star. 

We have approached the caracterisation of the magnetic properties of the star by performing a tomographic study. We do not detect any signature in the photospheric average line that is modulated by rotation (assuming the 0.88 day period). This non-detection at our noise level, together with the modulation of the light curve observed by \emph{Kepler}, imposes an upper limit to the cold spots that could be present at the surface of the star. We estimate that possible spots in the star occupy, at most, 0.002\% of the surface, a filling factor ten times larger than typical solar spots. 

Though spots have not been detected in the tomography, we still find them to be the most plausible explanation for the modulation of the light curve with a 0.88 day period detected by \emph{Kepler}. The main reason, apart from the fact that the amplitude is highly variable in time and that is fits very well with the width of spectral lines \citep{BoyajianWTF}, is that we find that the chromosphere of the star is much hotter than the temperature predicted by purely radiative losses. This means that the star has an active chromosphere. This fact is intrinsic to the star, as it is independent of the flux dimmings. This scenario is expected when the magnetic activity of the surface is present also in the outer layers of the atmosphere. In the Sun, the Network forms a filamentary bright pattern (in the core of strong chromospheric lines) all over the chromosphere. Though we can not rule out that other non-magnetic mechanisms (such as pulsations) can heat Boyajian's star chromosphere, we speculate that the magnetic activity can be responsible of the observed 0.88 day periodicity of the light curve and of the chromospheric heating. 

Some of the hypotheses found in the literature concerning the dimmings of the star are related to clouds of interstellar material or falling evaporating bodies (also referred as exocomets). Our spectroscopic survey does not strengthen either of the two scenarios. The spectral lines of the interstellar medium gas do not change during the dimmings. However, since we do not detect spectral signatures of the dust, we can not discard the interstellar dust as the star's occulter. At our sensitivity, we do not detect any absorption signature in the Ca\,{\sc ii} H and K lines due to exocometary tails. 

We compute the radial velocities of the average photospheric line in all our spectra. We detect, at a 1-2 sigma level, a drop in the radial velocity of the star during the \emph{Elsie} event. Since it has been shown that the dimming events are chromatic and likely due to optically thin dust \citep{BoyajianTEAM2018}, we claim that the dusty environment around the Boyajian's star has some clumps of thick material (with likely inclined and high impact parameter orbits) that produce the Rossiter-McLaughlin effect. However, since it is only a 1-2 sigma detection, we encourage future spectroscopic follow-up to strengthen or reject this scenario.

\section*{Acknowledgements}

The authors are grateful to D. Gandolfi for providing the FIES spectra published in Boyajian et al. (2015). Also, the authors want to acknowledge the support given from J. Telting with the data reduction of FIES spectra, and the support of the Mercator director allowing the extension of our observational campaign given the special circumstances. Financial support from the Spanish Ministry of Economy and Competitiveness through projects AYA2014-60833-P and AYA2014-60476-P are gratefully acknowledged. MJMG also acknowledges financial support through the Ram\'on y Cajal fellowship. GMK is supported by the Royal Society as a Royal Society University Research Support Fellow.

%%%%%%%%%%%%%%%%%%%%%%%%%%%%%%%%%%%%%%%%%%%%%%%%%%



% HSN: I had to change manually the Wright2016 reference. It crashes if more than 1 word in the second braces

\bibliographystyle{mnras}
%\bibliography{tabby} % if your bibtex file is called example.bib

% Don't change these lines
\bsp	% typesetting comment
\label{lastpage}
\end{document}